*Original Article*

# Variability Analysis of Isolated Intersections Through Case Study

Savithramma R M[1], R Sumathi[2], Sudhira H S[3]

[1,2]*Department of CSE, Siddaganga Institute of Technology, Karnataka, India.*
[3]*Gubbi Labs LLP, Karnataka, India.*

[1]savirmrl@sit.ac.in

*Abstract* - *Population and economic growth of urban areas have led to intensive use of private vehicles, thereby increasing traffic volume and congestion on roads. The traffic management in the city is a challenge for concerned authorities, and the signalized intersections are the primary interest of traffic management. Interpreting traffic patterns and current traffic signal operations can provide thorough insights to take appropriate actions. In this view, a comprehensive study is conducted at selected intersections from Tumakuru (tier-2 city), Karnataka, India. Data estimates traffic parameters such as saturation flow, composition, volume, and volume-to-capacity ratio. The statistical results currently confirm the stable traffic condition but do not ensure sustainability. The volume-to-capacity ratio is greater than 0.73 along three major arterial roads of study intersections, indicating congestion in the future as the traffic volume is increasing gradually, as per the Directorate of Urban Land Use and Transportation, Government of Karnataka. The statistical results obtained through the current study uphold the report. The empirical results showed 40% of green time wastage at one of the study intersections, which results in additional waiting delays, thereby increasing fuel consumption and emissions. The overall service level of the study intersections is of class C based on computed delay and volume-to-capacity ratio. The study suggests possible treatments for improving the service level at the intersection operations and sustaining the city's stable traffic condition. The study supports city traffic management authorities in identifying suitable treatment and implementing accordingly.*

## 1. Introduction

The role of transportation is vital in the development of any country. Urban areas are expanding with the rise in population. The economic and social affair department of the United Nations (Anon 2018) reported that around 55% of the population is city residents as of 2018, and it is expected to be 68% by 2050. The revised World Urbanization Prospectus (United Nations 2018) disclosed that the world's urban population is expected to be largely concentrated among a few countries. China, Nigeria and India together will account for 35% of the projected growth in urban areas between 2018 and 2050 across the world. The top ten cities in India recorded a growth rate of more than 20% in 1991-2001, with Surat growing at 87% (Balachandran, Adhvaryu, and Lokre 2006). It is expected that most of these cities will continue to grow at the same rate for the next decades. The study by (Duranton and Turner 2012) describes the direct relationship between population growth and the transportation system.

While the population has increased (Kohler 2012) (Sudhira and Gururaja 2012), public transport usage has deteriorated. As per a study (Group n.d.) conducted by the Louis Berger Group in Ahmedabad, one of the major cities in India, the passenger occupancy factor for the Ahmedabad Municipal Transport Service (AMTS) declined from 71% in the 1990s to 55% in 2000. One of the biggest problems of urban transport has been its inability to keep pace with the population growth in the city (Balachandran et al. 2006). The urban expansion (Sprawl) (Sudhira, Ramachandra, and Jagadish 2003) has adversely affected the sustainability of the transportation system (Zhao 2010). With the economic growth (Jones 2016), people gained the potential to own personal vehicles, thereby increasing the traffic volume, which is one of the common causes of congestion on roadways. Concerned urban management authorities are planning and allocating public funds to improve road infrastructure and other operational facilities to avoid congestion and ensure safety (Peixoto Neto et al. 2008).
The success of a good transportation system is in the number of people it can move from one place to another at an affordable cost, in reasonable comfort and at minimum cost to the environment. Under these circumstances, the traffic and transport studies support the development of project proposals focussing on transportation management in urban areas. Limited studies have been carried out for traffic characterization at intersections in the urban area (Pothula Sanyasi Naidu et al., 2015)(Shen and Wang, 2018). At the same time, a review of existing studies on intersection traffic modelling is presented (Pegu and Nath 2017).

Operations of signalized intersections are a part of the transportation management system. Junctions are the critical segments of the urban road network, while the delay is the prime concern of operational strategies implemented at the intersection. In this article, an attempt has been made to study the signalized intersections to assess their level of service. With this perspective, two junctions from Tumakuru, namely the Shivakumara Swamiji circle and the Town Hall circle, have been chosen for the current study. Tumakuru is a tier-2 city in India's southern state (Karnataka). The necessary data for the study is collected through videography. The parameters such as ratios of volume-to-capacity, traffic volume, traffic composition, saturation movements, signal configuration, delay, fuel consumption and emissions are analyzed to explore the operation level at selected city intersections.

In North America, the HCM (Highway Capacity Manual) TRB 2000 (Transportation Research Board, National Research Council, Washington, DC 2000) is the most widely used method for signalized intersections analysis. The intersection performance is defined in terms of mean delay. The delay is mapped against predefined service levels ranging from A to F, where 'A' represents the best Level of Service (LoS), and 'F' designates the worst service level operations. And the delay is computed as a function of many other factors, including signal configurations, temporal traffic variations, driver behaviour and environmental conditions. The analysis strategies defined by HCM are focused on uniform traffic composition. But the traffic conditions in India are highly heterogeneous. Hence, the analysis guidelines described in Indian Road Congress (IRC) (Indian Roads Congress 1990) and Indo-HCM (Year, Project, and Delhi 2010) are adopted in Indian scenarios.

The forthcoming sections of the article are laid as follows; the next section presents the available studies concerned with the current study. The area selected for the current study is introduced in section 3. The preliminary statistical analysis conducted over signal operations and the data collection concerning traffic at selected intersections are presented in sections 4 and 5, respectively. The results of variability analysis concerned with different parameters are presented in section 6, and the corresponding observations and probable solutions are discussed in section 7. Finally, the paper concludes with the report of the proposed study.

## 2. Literature

By identifying the delay variability, more reliable traffic signal configurations can be estimated to improve the service level (Transportation Research Board, National Research Council, Washington, DC 2000). Therefore, many researchers worldwide have presented studies on intersection operations and traffic conditions in specific regions. The analysis of signalized intersections in Waterloo and Kitchener cities of Canada is presented in (Hellinga and Abdy 2008) with the objective of day-to-day peak-hour traffic volume variability implications on delay. The authors (Darma and Karim 2005) researched to determine the set of components (variables) of the HCM delay model that influences control delay using the capacity softwares Transyt-7F and SIDRA.

The research is in progress to develop advanced models for delay estimation. In this connection, many authors worldwide have already presented the models. Different analytical models (Fu and Hellinga 2000) (Akgungor and Bullen 2007) (Chen et al. 2013) (Fawaz and El Khoury 2016) were presented to estimate delay. In contrast, the studies including (Amrutsamanvar and Arkatkar 2018) and (Yesufu et al. 2019), investigated the contribution of delay at signalized intersections on overall travel time variability along the route. The variance of overall delay at the signalized intersection is estimated by (Fu and Hellinga 2000) based on delay evolution patterns in oversaturated and undersaturated traffic situations. But, (Akgungor and Bullen 2007) consider the traffic flow variations to compute the delay. The model involves a delay parameter k, expressed as a function of the degree of saturation. The authors have proved that the proposed model performs well in all expected traffic conditions.

Delay estimation at the pre-timed signal-operated intersection is presented in (Chen et al. 2013) by considering traffic arrival distributions using traditional cumulative curves. A uniform control delay is modelled for undersaturated intersections (Fawaz and El Khoury 2016). The delay model (Webster 1958) proposed is appropriate for homogeneous traffic with lane-adhered traffic situations. Hence, its modified versions were presented by (Hoque and Imran 2007) (Minh et al., 2010) (Preethi, Varghese, and Ashalatha 2016) (and Saha, Chandra, and Ghosh 2017) to suit the uniform traffic conditions in India.

A comparison of various delay models like deterministic queuing, shock wave theory-based, Webster, HCM, Australian Capacity Guide, etc. has been presented (Dion, Rakha, and Kang 2004). The delay estimates were analyzed under high and low traffic conditions and observed that all models work better in case of low traffic demand while showing differences in case of saturated conditions. Performance of the traffic signal control system is expressed in terms of delay encountered by each vehicle in a waiting queue. Authors (Ghavami, Kar, and Ukkusuri 2012) have compared the delays caused by three different traffic signal control algorithms, namely, static (state-independent) Fixed-Time Scheduling (FTS), Dynamic Maximum Weight (backlog) Scheduling (DMWS) algorithm and Adaptive length MWS (AMWS) algorithm. A study has been presented by (Feng et al. 2014) focusing on the impact of the actuated signal operations to manage intersection in terms of waiting delay.

Reliable estimation of traffic saturation flow is crucial for proper intersection design and operations. Hence, the authors (Chodur, Ostrowski, and Tracz, 2016) presented a comparative analysis of saturation flow estimates at urban and rural intersections. The saturation flow estimation model is proposed by (Saha, Chandra, and Ghosh 2018) for signalized intersections under unstructured traffic scenarios. Variability analysis of saturation flow estimation is given by (Nguyen 2016) by considering cities with heterogeneous traffic but mostly depend on two-wheelers as a case study. The methodology has been described in (Chand, Gupta, and Velmurugan 2017) (and Marfani and Dave 2016) for computing saturation flow and Passenger Car Unit (PCU) under heterogeneous traffic. The authors (Vasantha Kumar et al., 2018) explored the impact of urbanization on traffic flow rate at junctions through the study by considering a three-legged intersection from Vellore, India, as a case study.

The relationship between the volume-to-capacity ratio and the service level at junctions under traffic heterogeneity is discussed in (Othayoth and Rao 2020). In contrast, the intersection operation level in South Africa is evaluated by (Bester and Meyers 2007) based on saturation flow rate. A study was carried out by (Prasanna Kumar and Dhinakaran 2013) to characterize the traffic patterns to assess the level of service, while (Qu et al. 2013) (A J Mavani et al. 2016) conducted a qualitative and quantitative analysis of traffic discharge flow rate to know the dispersion characteristics of traffic flow that helps in framing the reliable theoretical basis for designing the traffic signal plans.

An appropriate mathematical model to express commuters' behaviour helps minimize congestion, travel time, and fuel consumption, thereby mitigating environmental pollution. In this direction, (Macioszek and Iwanowicz 2021) proposed a model to estimate the maximum queue size at the intersections based on driver behaviour. The traffic queuing behaviour has been analyzed by (Verma et al. 2018), and they proposed a modified Webster delay model to optimize the traffic signal plan. The study presented by (Saw, Katti, and Joshi 2018) mainly concentrated on various traffic movement patterns during queuing and discharge periods which causes travel delay and (Albrka Ali, Reşatoğlua, and Tozan 2018) analyzed the traffic flow rate at roundabouts using SIDRA-5 software.

A study presented in (Darshan Patel, P. N. Patel 2018) mainly focused on the effect of parameters such as traffic arrival rate, discharge rate and composition on saturation flow rate and road capacity at the signalized intersection. An algorithm for traffic scheduling is proposed by (Roopa et al. 2020) using the Internet of Things (IoT) to avoid traffic jams, thereby enhancing the traffic throughput at signalized junctions and the dynamic use of left lanes to improve the traffic flow rate along through lanes is proposed in (Zheng et al. 2020).

## 3. Study area

The literature review discussed in the previous section unveils that very few studies have been conducted on urban signalized intersections in India. In particular, studies on smart cities are rarely carried out concerning signalized intersections. With this preliminary investigation, Tumakuru, a tier-2 city in Karnataka state (*Figure 1*), has been chosen for the study. A comprehensive analysis has been conducted in Tumakuru, selected under Smart City Mission (SCM) as an initial step. The SCM was launched by the Indian government on 25 June 2015. The main objective of this mission is to promote complete and sustainable cities able to provide their civilians with a clean environment and decent quality of life by applying smart solutions. In this direction, the focus of the mission is to develop creative and replicable models.

Currently, the Smart City Project (SCP) is under progress in Tumakuru, and six junctions along the BH (Bengaluru Honnavara) road are operated with an automated signal system (Banerjee 1971). Among six, two isolated signalized intersections are selected for the study: Shivakumara Swamiji Circle (SSC) and Town Hall circle (THC). The study circles are located 2.1 km away, with the Bhadramma circle reclining. The approaching roads have one to three lanes with one-way traffic movement with no pressure from roadside parking, bus stops, or any other resistance to the traffic flow.

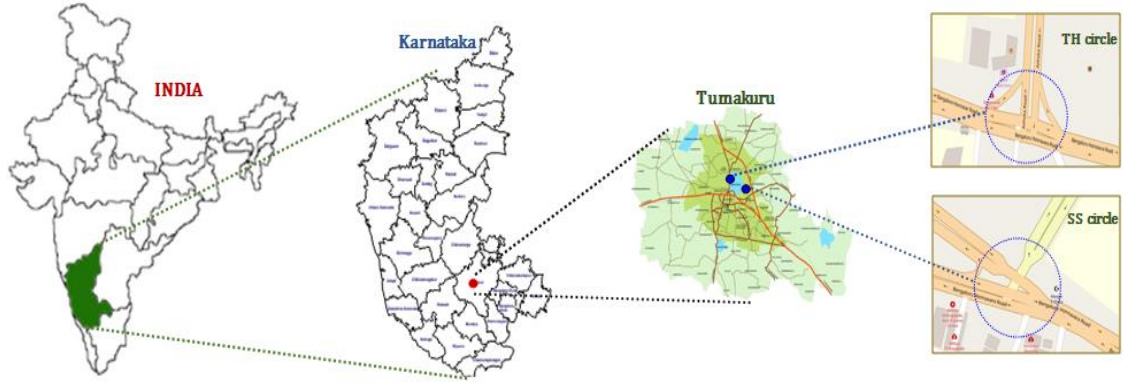

**Fig. 1 Location of the study area**

### 3.1. Shivakumara Swamiji circle

It is a five-legged junction with two minor and three major roads intersecting. In the context of this paper, major road indicates heavy traffic flow, whereas minor road indicates lower traffic flow. The roads are approaching from Sira-gate (*SR1*), Bhadramma circle (*SR2*), Adithya hospital (*SR3*), Batawadi (*SR4*) and Stadium (*SR5*). SR1, SR2 and SR4 are major roads, while SR3 and SR5 are minor roads. In a phase, the traffic can move in all possible directions from an approaching road. SR1 and SR5 are permitted with a free left. The top view of approaching roads of the SS circle is presented in *Figure 2*.

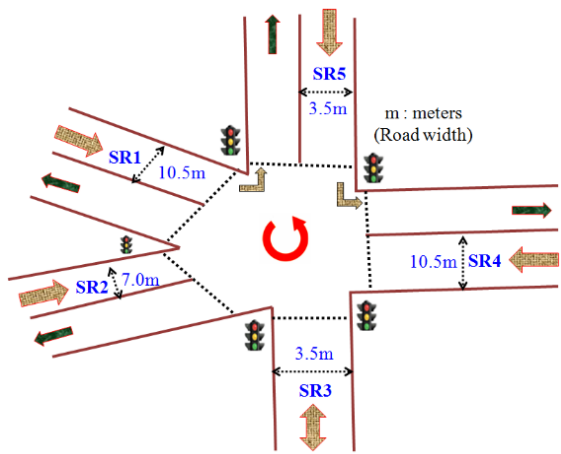

**Fig. 2 Shivakumara Swamiji circle**

### 3.2. Town Hall circle

It is a cross (four-legged) intersection with one minor and three major roads crossover. The intersecting roads are approaching from Ashokaroad (*TR1*), Call-tax circle (*TR2*), Municipal corporation (*TR3*) and Bhadramma circle (*TR4*). TR1, TR2 and TR4 are major roads, while TR3 is a minor road. The direction of traffic movement is allowed from one road to all other roads in one phase. TR1 and TR2 are permitted with a free left. The top view of intersecting roads of the TH circle is presented in *Figure 3*.

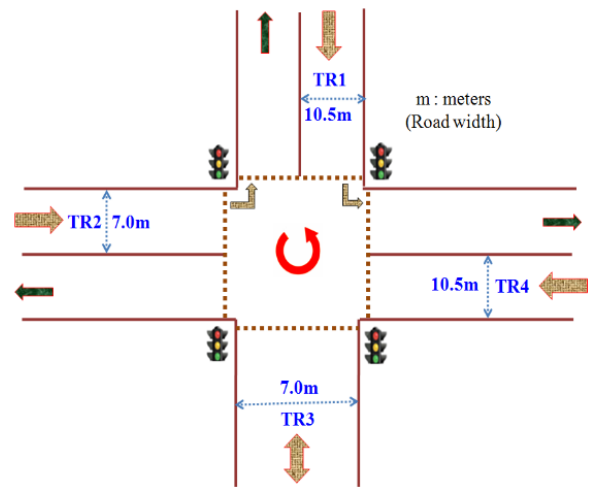

**Fig. 3 Town Hall circle**

## 4. Preliminary analysis

The initial analysis has been carried out to recognize the probable peak hour/s in a day and heavy-traffic-flow-day/s in a week in the city by considering two intersections introduced in the previous section. This initial analysis is based on currently active signal operations. The statistical process is applied over aggregates of cycle lengths utilized throughout the day from morning 8:00 to evening 9:00. The day-wise statistics are given in *figures 4(a)* and *5(a)* for the SSC and THC, respectively. The x-axis represents the time-slices of 30 minutes, whereas average cycle lengths in seconds are presented on the y-axis. The statistics demonstrated that the average cycle lengths are longer during weekdays than weekends (Saturday and Sunday). However, the average cycle length on Saturdays lies between weekdays and Sundays.

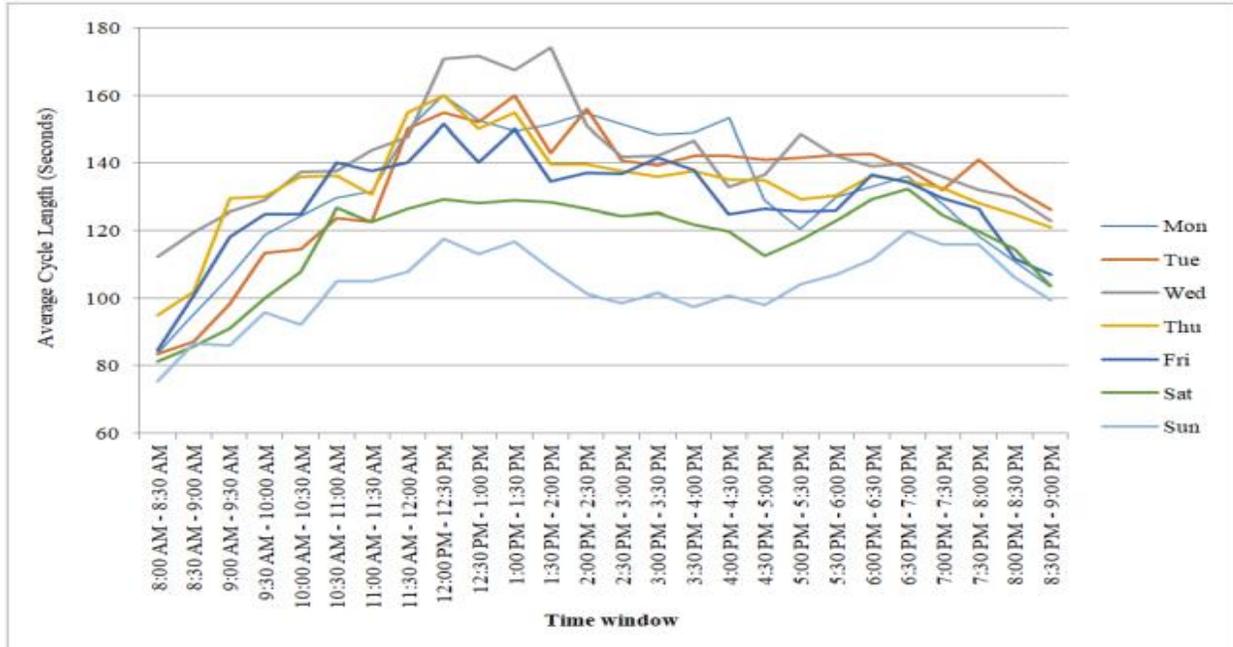

**Fig. 4(a) Day-wise average cycle length at SS circle**

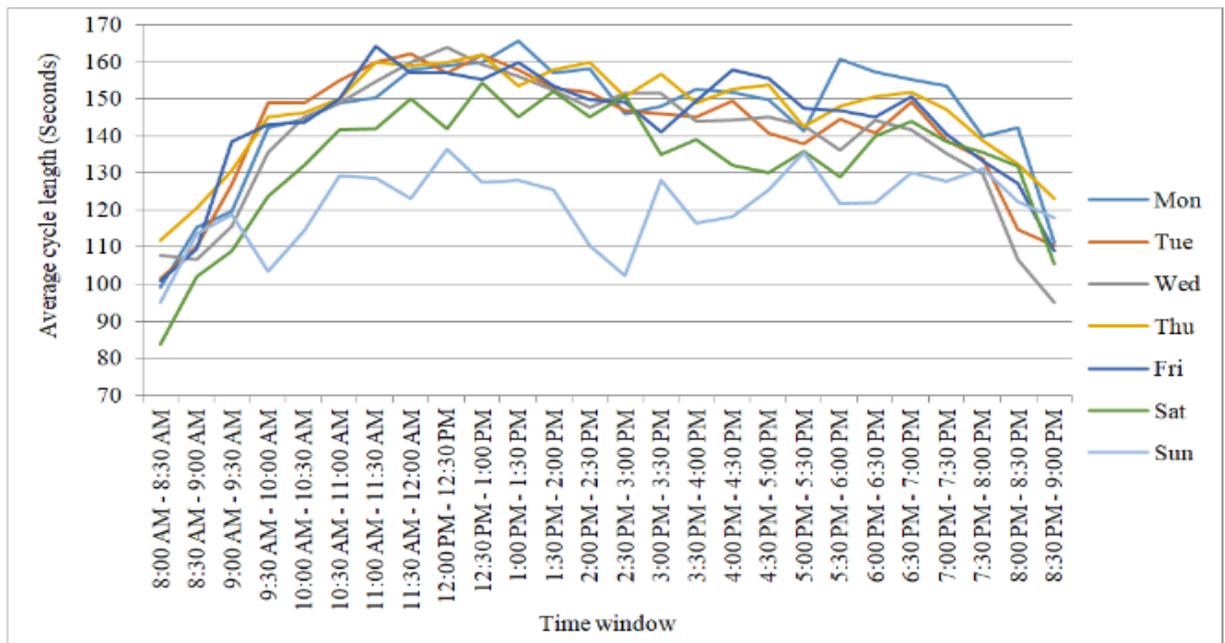

**Fig. 4(b) Day-wise average cycle length at THC**

With this prior information, only weekdays are considered for computing average cycle lengths in a day to know the peak-hour/s. The average cycle lengths utilized at SSC and THC over the half-an-hour window are depicted in Figures *4(b)* and *5(b),* respectively. The time slices are given on the x-axis, while the average cycle length in seconds is plotted along the y-axis. The time window of 11:00 AM to 1:00 PM was found to be a peak hour in a day at study intersections. Further, these observations are a base for additional data collection at selected intersections.

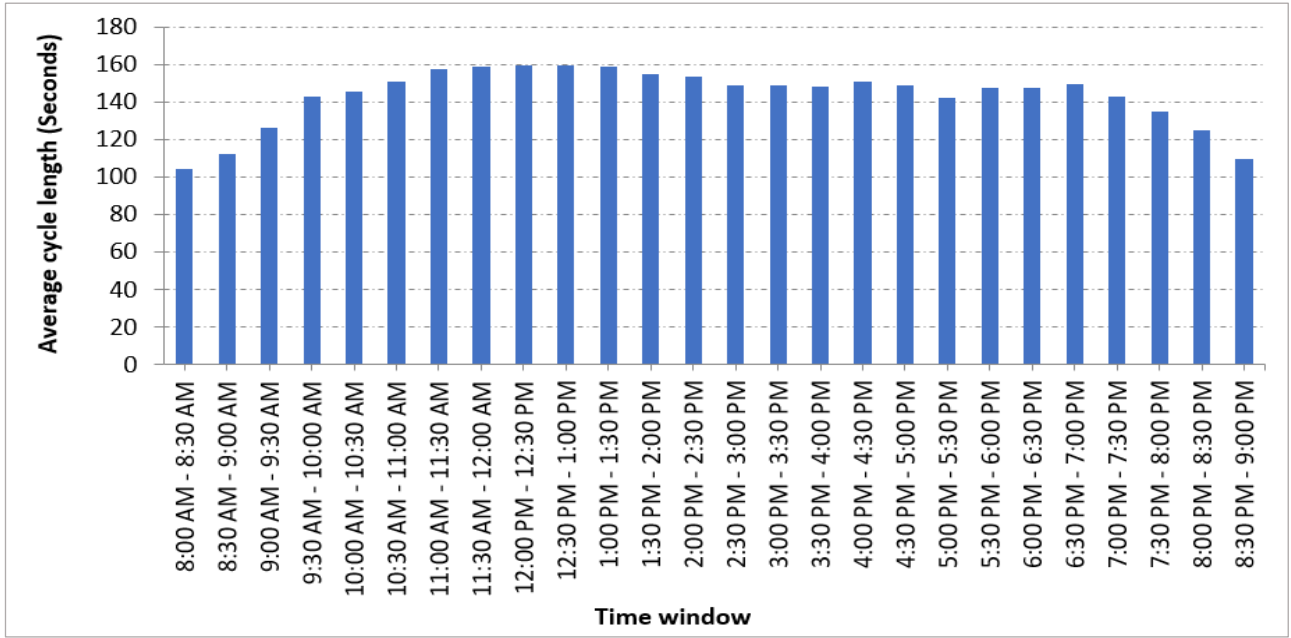

**Fig. 5(a) Average cycle lengths at SSC**

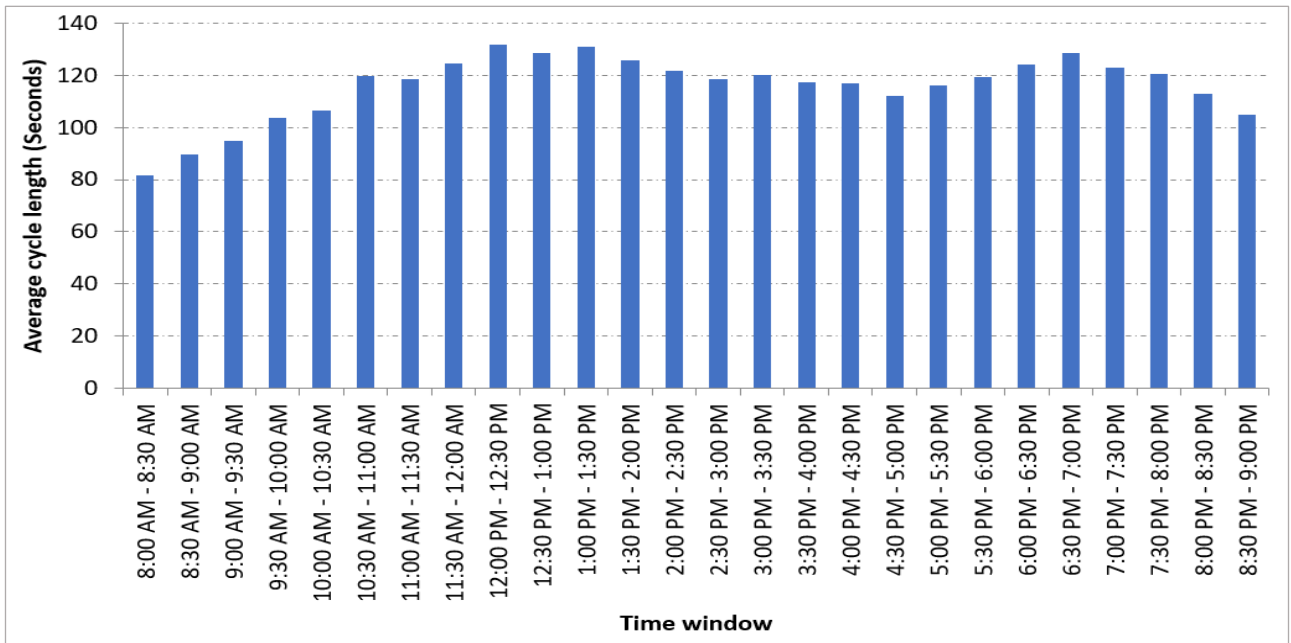

**Fig. 5(b) Average cycle lengths at THC**

## 5. Data collection

The effectiveness of an Intelligent Transportation System (ITS) is mainly relying on the tools and technology used to collect and access the data from real-time scenarios. In recent decades sensor technology has become ubiquitous and largely applied across most sectors, and Traffic Management systems (TMS) is one among them. Seamless integration of sensors with vehicles and road infrastructure can result in efficient TMS development. With the effort of researchers, a wide range of sensors are available and are discussed in s(Ashwini and Sumathi 2020). Cameras are widely used sensors to monitor traffic conditions at intersections. Currently, each signalized intersection in Tumakuru city is equipped with high-definition cameras to capture traffic patterns. The data concerning traffic patterns and signal control operations at each approaching road are collected through videography.

As the traffic in Tumakuru is highly heterogeneous, the overall traffic is classified as two-wheelers, auto-rickshaws, cars, Light Commercial Vehicles (LCVs) and Buses. Then, the collected traffic data is normalized to exhibit uniform traffic by converting the vehicle count into PCU (Passenger Car Unit). PCU is the standard unit representing the uniform traffic structure where the different types of automobiles are transformed into cars. The ICR-106-1990 (Indian Roads Congress 1990) guidelines are used in this article for conversion to PCU. The recommended equivalent PCU factors are presented in *Table 1*.

**Table 1. PCU factor and type of vehicles**

| Sl. No. | Type of Vehicle | Equivalent PCU factors | |
|---|---|---|---|
| | | Traffic composition (%) | |
| | | <5% | >5% |
| 1 | Two-wheelers | 0.50 | 0.75 |
| 2 | Cars | 1.00 | 1.00 |
| 3 | Auto rickshaws | 1.20 | 2.00 |
| 4 | LCV | 1.40 | 2.00 |
| 5 | Trucks/Buses | 2.20 | 3.70 |

Based on the preliminary statistics, the data is collected weekly during peak hours (from 11:00 AM till 1:00 PM). The vehicle counts and signal operation details are recorded manually by watching the videos. The recorded data is categorized as signal data and traffic data. *Table 2* presents the aggregates of collected data.

**Table 2. Data collected accumulated from the study intersections**

| Road ID | Cycle length (Sec) | Red time (Sec) | Green time (Sec) | Two-wheelers (PCU) | Autorickshaws (PCU) | Cars (PCU) | Light Commercial Vehicles (PCU) | Bus (PCU) |
|---|---|---|---|---|---|---|---|---|
| SR1 | 152 | 120 | 32 | 23 | 10 | 12 | 4 | 2 |
| SR2 | 152 | 117 | 35 | 33 | 30 | 10 | 1 | 7 |
| SR3 | 152 | 140 | 12 | 6 | 2 | 2 | 1 | 0 |
| SR4 | 152 | 107 | 45 | 35 | 28 | 18 | 4 | 4 |
| SR5 | 152 | 124 | 28 | 8 | 4 | 10 | 1 | 2 |
| TR1 | 119 | 94 | 25 | 12 | 14 | 4 | 4 | 0 |
| TR2 | 119 | 79 | 40 | 30 | 16 | 7 | 6 | 8 |
| TR3 | 119 | 110 | 9 | 7 | 1 | 3 | 4 | 0 |
| TR4 | 119 | 74 | 45 | 38 | 25 | 9 | 7 | 8 |

## 6. Variability analysis

A *Z-test* is performed to explore the spatial-temporal variations in the data to understand whether both intersections exhibit the same or unique traffic behaviour. The *p-value* computed around the *means* is less than 0.0001, signifying the unique traffic inflow behaviour at both intersections. The magnitude of disparity in *means* of data points is characterized by a box plot, as depicted in *Figure 6*.

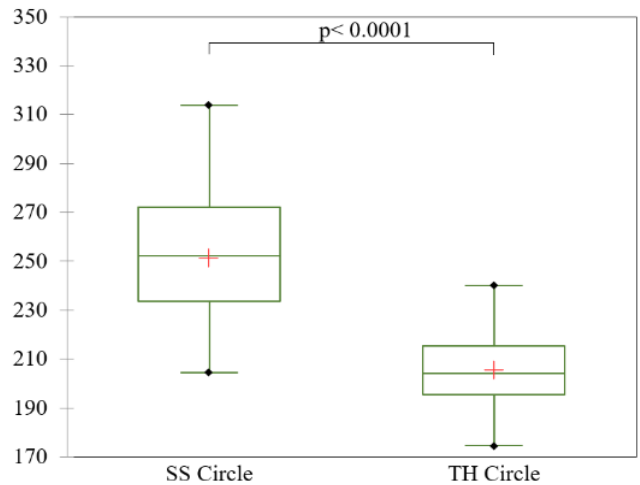

**Fig. 6 Box plot of z-test conducted over traffic inflow**

Further, the traffic inflow behaviour at each approaching road of individual intersection is compared against each other through a *z-test*. The p-values computed for each pair are given in *Table 3(a) and 3(b)* for SSC and THC, respectively. The results showed each approaching road's unique traffic inflow behaviour at both intersections. Hence, with this primary inference, each approaching road is evaluated individually. Various parameters concerning signal operations and traffic are aggregated over peak hours (Temporal) at each road of study intersections (Spatial).

Table 3(a): P-Values for each pair of approaching roads at SSC

| SR2 | 0.0002 | | | |
|---|---|---|---|---|
| SR3 | 4.16E-18 | 3.77E-29 | | |
| SR4 | 0.113 | 0.028343 | 1.28E-22 | |
| SR4 | 7.53E-12 | 1.91E-22 | 0.024383 | 6.23E-16 |
| | SR1 | SR2 | SR3 | SR4 |

Table 3(b). P-Values for each pair of approaching roads at SSC

| TH2 | 0.001227 | | |
|---|---|---|---|
| TH3 | 0.26406 | 1.73E-05 | |
| TH4 | 7.05E-08 | 0.022172 | 1.71E-10 |
| | TH1 | TH2 | TH3 |

### 6.1. Traffic volume

The primary factor driving the signal controller is the traffic volume. Which in turn influences the waiting delay of the vehicle. The peak-hour traffic volume in PCU crossed the SS circle and TH circle in an hour is given in *Figures 7(a)* and *7(b)*, respectively. The approaching roads are specified along the x-axis, and the y-axis represents the volume in PCU/hr.

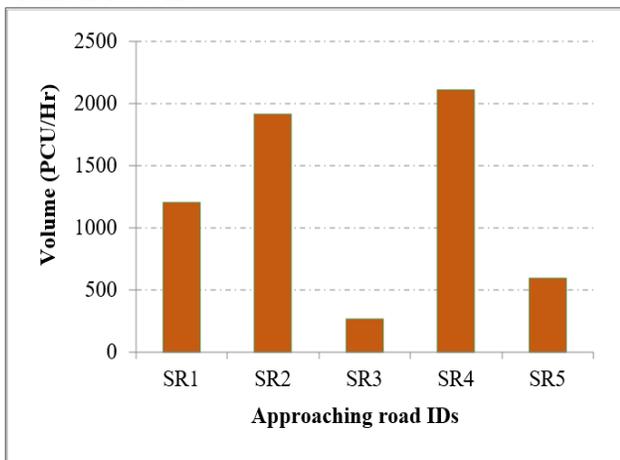

Fig. 7(a) Traffic volume in PCU at SSC

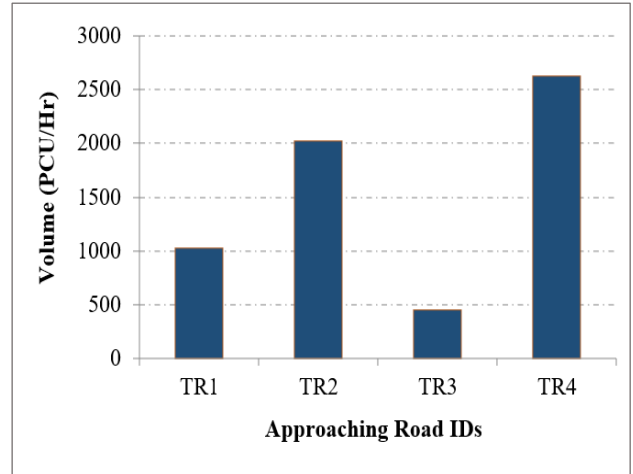

Fig. 7(b) Traffic volume in PCU at THC

### 6.2. Traffic composition

The traffic classification is a part of an advanced traffic management system used to monitor traffic flow, vehicle parking, intersection operations and safety enforcement. The traffic classification-related review is presented in (Shokravi et al. 2020). The traffic signal operations at intersections are greatly affected by the traffic composition. As the traffic condition in the city is highly heterogeneous, the traffic structure is categorized into five classes for simplification (Gowri and Sivanandan 2008). The classification includes auto-rickshaws, two-wheelers, cars, buses and Light Commercial Vehicles (LCVs). The percentage of each class of vehicle (traffic composition) crossing the study intersections during peak hours is provided in *Figures 8(a)* and 8*(b)*.

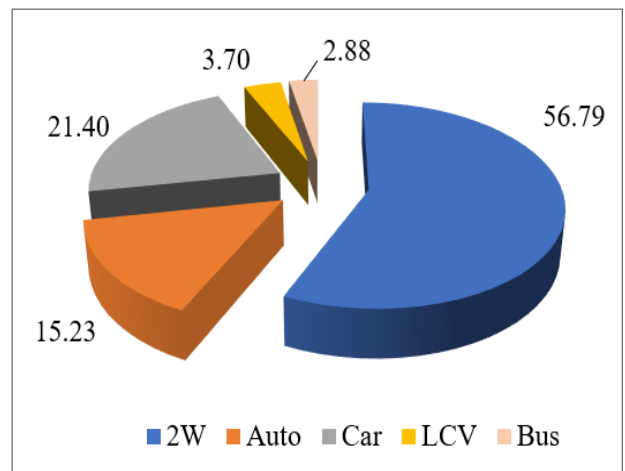

Fig. 8(a) Traffic composition in % at SSC

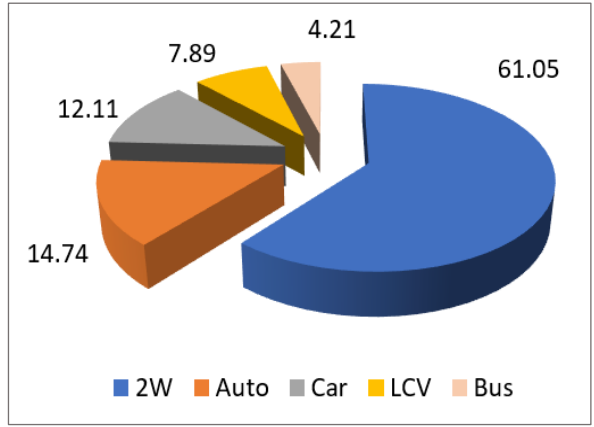

**Fig. 8(b) Traffic composition in % at THC**

### 6.3. Ratios of Volume-to-capacity

The ratios of volume versus capacity are used to express the level of congestion at a specific road segment (Regional, Commission, and Program 2007). It is one of the decisive parameters when assessing the sustainability of the existing signal controller. Capacity is the maximum flow rate at which traffic passes through a particular road segment in an hour under general conditions (Chandler et al., 2013). 'C' is the capacity of each approaching road measured based on ICR-106-1990 (Indian Roads Congress 1990), whereas 'V' is the observed peak-hour volume. The volume-to-capacity ratios *(V/C)* of each approaching road are computed and given in *Table 4*.

**Table 4. Volume-to-capacity ratios at study intersections**

| | RID | Number of Lanes | Capacity Volume in PCU/Hr. (As per IRC-106) (C) | Observed Volume in PCU/Hr. (V) | V/C |
|---|---|---|---|---|---|
| SS circle | SR1 | 3 (One-way) | 3600 | 1206 | 0.34 |
| | SR2 | 2 (One-way) | 2400 | 1918 | 0.80 |
| | SR3 | 1 (Two-way) | 2400 | 270 | 0.11 |
| | SR4 | 3 (One-way) | 3600 | 2110 | 0.59 |
| | SR5 | 1 (One-way | 1500 | 594 | 0.40 |
| TH circle | TR1 | 3 (One-way) | 3600 | 1029 | 0.29 |
| | TR2 | 2 (One-way) | 2400 | 2027 | 0.84 |
| | TR3 | 2 (One-way) | 2400 | 454 | 0.19 |
| | TR4 | 3 (One-way) | 3600 | 2632 | 0.73 |

### 6.4. Saturation flow

The traffic Saturation Flow (SF) is a crucial and macro performance metric used to assess the operations at the signalized junction (Bester and Meyers 2007). The potential capacity of an intersection is indicated with saturation flow with ideal conditions as defined in TRB 2000. SF is the count of vehicles exiting from a junction during an effective green hour. At the same time, the effective green time is a part of the allocated green length during which maximum possible traffic departure takes place. The standard mathematical model for SF computation in terms of PCU is given in equation (1) (Nguyen 2016) (Saha et al. 2018).

$$S = \frac{N}{g_e} * 3600 \qquad (1)$$

Where the term *S* indicates the SF (PCU/hour) at an intersection along a particular direction of traffic discharge, *N* is the vehicle count (PCU) exiting the junction ineffective green time interval $g_e$(Seconds).
As per IRC-106-1990 (Indian Roads Congress 1990), the saturation flow model is given in equation (2).

$$S = 525 * W \qquad (2)$$

The width of an approaching road is represented as W and measured in meters. The saturation flow estimated according to equations (1) and (2) is represented in *Table 5*.

**Table 5. Saturation flow**

| Intersection | RID | Road width in meters (W) | Effective green in seconds ($g_e$) | PCUs exited during $g_e$ (N) | SF1: saturation flow as per eq. (1) (PCU/Hr.) | SF2: saturation flow as per eq. (2) (PCU/Hr.) | Difference (SF1 – SF2) |
|---|---|---|---|---|---|---|---|
| SS circle | SR1 | 10.5 | 24 | 48 | 7200 | 5513 | 1688 |
| | SR2 | 7.0 | 31 | 77 | 8942 | 3675 | 5267 |
| | SR3 | 3.5 | 10 | 9 | 3240 | 1838 | 1403 |
| | SR4 | 10.5 | 35 | 83 | 8537 | 5513 | 3025 |
| | SR5 | 3.5 | 16 | 21 | 4725 | 1838 | 2888 |
| TH circle | TR1 | 10.5 | 18 | 30 | 6000 | 5513 | 488 |
| | TR2 | 7.0 | 35 | 61 | 6274 | 3675 | 2599 |
| | TR3 | 7.0 | 9 | 17 | 6800 | 3675 | 3125 |
| | TR4 | 10.5 | 38 | 82 | 7768 | 5513 | 2256 |

## 6.5. Green time splits

Currently, the traffic signal planner operates adaptively in response to the traffic volume at study intersections. The average cycle lengths allocated at the SS and TH circles are 152 and 119 seconds, respectively. The empirical results of processed data confirm the time-wise and day-wise variability in signal configuration. The percentage of operational green splits utilized at the SS circle and TH circle is depicted in *Figures 9(a)* and *9(b)*, respectively.

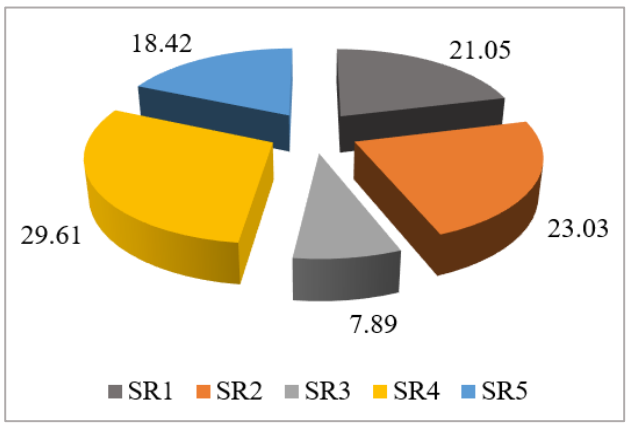

**Fig. 9(a) Percentage of green splits at SSC**

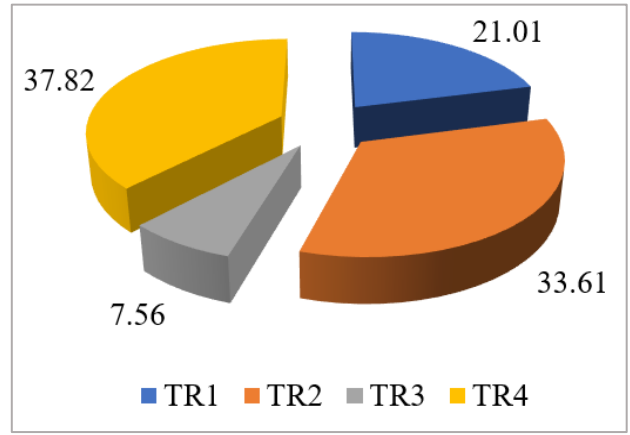

**Fig. 9(b) Percentage of green splits at THC**

## 6.6. Green time utilization

Efficient utilization of allocated green time will enhance the overall traffic discharge rate. In other words, the potentiality of signal operations lies in the accurate allocation of green time per traffic demand in each direction. There is a probability of green time wastage even with the adaptive traffic controller at the time due to erroneous data detected from the environment. In this view, the signal operations at study intersections are assessed based on the number of vehicles discharged and the corresponding green time allocated. The empirical results are presented in *Figures 10(a)* and *10(b)* at the SS and TH circles, respectively.

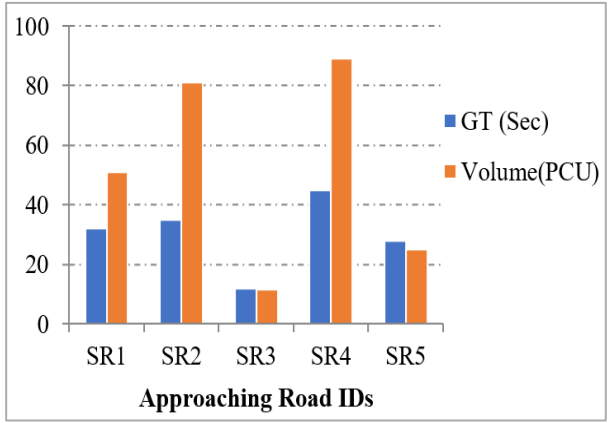

Fig. 10(a) Allocated green time v/s PCU at SSC

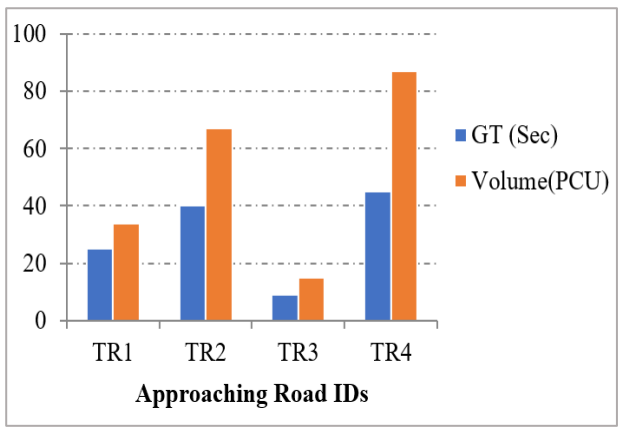

Fig. 10(b) Allocated green time v/s PCU at THC

*6.7. Vehicle waiting delay*

Travel delay is the unwanted time experienced by the traveler along the journey in a road network. However, the majority of delay is encountered at the intersection. The primary objective of optimizing traffic signal controllers is to mitigate the overall waiting delay at intersections. The time spent in traffic queues at the intersection is termed control delay. Delay is an important performance measure of an intersection. Several delay estimation models are proposed by various authors and researchers worldwide. In this paper, the average control delay is estimated using the modified Webster method as given in equation (3), proposed by (Saha et al. 2017) based on Indian scenarios.

$$d_c = 6.23 + \frac{0.5*C*(1-\frac{g}{C})^2}{1-X*\frac{g}{C}} - 15.35 * R_p \quad (3)$$

Where $d_c$ represents control delay (seconds), C indicates cycle length (seconds), the green time (seconds) is denoted by G, and X represents the ratio of volume-to-capacity (v/c), v is the volume (PCU), c is the capacity of the approaching road in PCU and $R_p$ is plathoon ration as given in equation (4).

$$R_p = \frac{PVG}{PTG} \quad (4)$$

Where the % of traffic arriving in the green period is PVG, whereas the % of the overall green period in a cycle is PTG, the estimated delay at each road of study intersections is presented in *Table 6.*

Table 6. Average control delay computed at SS circle and TH circle

| SS circle | | | TH circle | |
|---|---|---|---|---|
| RID | Delay (Seconds) | | RID | Delay (Seconds) |
| SR1 | 50.24 | | TR1 | 42.44 |
| SR2 | 56.42 | | TR2 | 37.52 |
| SR3 | 60.51 | | TR3 | 45.32 |
| SR4 | 52.76 | | TR4 | 34.02 |
| SR5 | 60.46 | | | |

## 7. Discussion

The key findings of the current study are discussed in this section. The current study objective is to assess the probable Level of Service (LoS) at study intersections. The LoS is a performance measure used to express the signal operational quality of any intersection. The factors like speed, travel time, travel delay, free mobility and safety are the deciding elements of service levels. The level of service is designated with the range 'A' to 'F', where 'A' denotes the best operation level, and 'F' represents the worst operating condition. HCM 2000 (Transportation Research Board, National Research Council, Washington, DC 2000) has defined several methods to compute the LoS for junctions and other roadways.

This article uses delay and volume-to-capacity ratio methods to assess the LoS. The delay ranges recommended by HCM to assess the LoS at an intersection are based on uniform traffic structure, which might not suit Indian traffic scenarios. Hence, the authors (Saha, Chandra, and Ghosh, 2019) have studied Indian traffic scenarios and derived delay ranges suitable for heterogeneous traffic conditions. The delay range recommended by HCM 2000 and (Saha et al. 2019) is given in *Table 7*. The volume-to-capacity ratio ranges described in (Washington 2005) are presented in *Table 8*.

**Table 7. Recommended delay ranges for LoS assessment**

| LoS | A | B | C | D | E | F |
|---|---|---|---|---|---|---|
| By (Saha et al. 2019) (Sec/Vehicle) | ≤10 | 10 - 45 | 45 - 65 | 65 - 100 | 100 - 135 | >135 |
| By HCM (Sec/Vehicle) | ≤10 | 10 - 20 | 20 - 35 | 35 - 55 | 55 - 80 | >80 |

**Table 8. Recommended volume-to-capacity ranges for LoS assessment**

| LoS | F | E | D | C | B | A |
|---|---|---|---|---|---|---|
| V/C | ≥1 | 0.90 - 0.9 | 0.80 - 0.89 | 0.70 - 0.79 | 0.60 - 0.69 | <0.60 |

According to the estimated delays in *Table 6,* the average delay is 56 seconds and 36 seconds at the SS circle and TH circle, respectively. The major arterial roads, including SR2 and SR4 from the SS circle and TH2 and TH4 from the TH circle, are considered for computing the average delay. As per the standard delay ranges provided in *Table 7*, the level of service at the TH circle and SS circle is class 'B' and 'C', respectively, based on (Saha et al., 2019). But, as per HCM standards, the LoS at the TH circle is class 'D 'and LoS at the SS circle is class 'E'. The operation level at both junctions is class 'C' as per *Table 8* based on the volume-to-capacity ratio. At the macro level, the study ascertains the probable operational conditions at signalized intersections in Tumakuru and similar cities in India.

Several other factors are considered while deriving the service levels for the intersections. The traffic volume is higher (≈70%) along the major arterial roads as they are connected with major cities and commercial areas. The two-wheelers share the larger percentage of overall traffic composition while buses share the minimum. 59% of two-wheelers, ≈17% of cars, and ≈4% of buses are found in one cycle. These statistics express the extent of private vehicle usage. Also, the volume-to-capacity ratios computed indicate the probability of congestion in the future and raise the question of the sustainability of current operational strategies. As per (Regional et al. 2007), the volume-to-capacity ratios greater than or equal to one indicate congestion, and less than one indicates no congestion. Hence, current traffic conditions and corresponding operational strategies are stable but do not merely ensure longer-term sustainability. The volume-to-capacity ratio is 0.8, 0.84, and 0.73 at SR2, TH2 and TH4, respectively, which is heading towards congestion shortly as the traffic is increasing gradually as per the report of the year 2014 by DULT (Directorate of Urban Land Transport, Karnataka Government). All these empirical statistics demand the treatment of operational traffic systems at signalized intersections.

The saturation flow is prime important in designing and controlling signalized junctions. The current saturation flow rate at study intersections is above the standard values as per IRC-106-1990 (Indian Roads Congress 1990). As the saturation flow influences signal operations, it is necessary to analyze the signal controller that is currently active in the city. The larger part of the cycle time is allocated for major arterial roads compared to other sub-arterial approaching roads. 71.43% of the green time is utilized by TH2 and TH4 together at the TH circle, whereas 56.64% of the green time is allocated for SR2 and SR4 at the SS circle. However, empirical results showed that the green time allocated at the SSC for SR3 and SR5 is underutilized. Waste of intersection's green time is one of the major issues to be addressed to mitigate the delay. The green time allocated versus the number of vehicles that crossed the intersection during the allocated green time is plotted in *Figure 10*. It is noticed that the green time (Sec) to volume (PCU) ratio is generally less than one at all approaching roads except SR3 and SR5, indicating suboptimal allocation of green time, i.e. approximately 35% to 40% of the green time is wasted.

There are several ways to treat intersections (FHWA 2004). Optimizing traffic signal operations is the first probable treatment to mitigate the green time wastage and enhance the saturation flow. With this view, various solutions exist to implement optimized traffic signal control systems (Savithramma and Sumathi 2020). The most important issue in tier-2 cities is the intensive use of private vehicles (DULT report 2014). The primary solution to overcome this problem is to shift the mode of transportation of commuters to the Public Transit System (PTS). In this direction, PTS was launched in Tumakuru in 2011, and the city transportation services were assessed for 2012 (Transport 2013). The evaluators observed a significant modal shift of 17% and a decline in two-wheeler usage after the introduction of PTS. The report declared 69% and 60% of two-wheelers usage before and after respectively. The current study observed approximately 58% of two-wheelers at study intersections indicating a mere reduction in two-wheelers usage.

The observed statistics imply that attracting commuters to public transport is challenging. However, public transportation usage may be encouraged through various incentives and offers like bas fare reduction or discounts, permitting passes, employer-provided subsidies or partial payments by the employer and reduction in pre-tax payroll. Conduct awareness programs for college students to encourage them to opt for a walk or to use bicycles (US

Dept. of Transportation). Implementation of an advanced passenger information system, allocating dedicated lanes in the road network for PTS and prioritizing the buses at signalized intersections to ensure reliability can also promote the use of PTS. Carry-through of ideas mentioned above and policies with public participation increases the likelihood of reflecting public needs inaction taken by the public agencies or authorities. Still, the public participatory process (NUTP, Ministry of Urban Development 2014) is challenging.

An alternative solution to reduce the number of private vehicles is to adopt the lottery system to purchase a vehicle. It is a way of restricting vehicle ownership. According to the Vehicle Quota System (VQS) or License Plate Lottery Policy (LPLP), the number of vehicle registration in a particular region is restricted to some threshold. Hence, they have to win a licensed lottery to own a car. Singapore successfully reduces private vehicles, thereby mitigating congestion by introducing VQS (Koh 2003). Also, the same system was implemented in Beijing under the name LPLP in 2011(Yang et al. 2014) to reduce traffic volume, air pollution and congestion (Zhu, Du, and Zhang 2013).

Environmental pollution is the prime concern of traffic management. The petroleum products utilized by vehicles produce greenhouse gases ($CO_2$, Na) which are highly toxic. The vehicle not only emits gases during mobility but also in its idle condition. Idling is common in traffic congestion and at signalized intersections. A study by (Bhandari, Parida, and Singh 2013) has presented the amount of fuel consumed by various vehicles in an idle state. For simplification, it is assumed that all vehicles remain in an idle state during the waiting period at signalized intersections. The experimental values concerning fuel consumption given in (Bhandari et al. 2013) are referred to for computing fuel consumption. The amount of gas emitted per liter of fuel consumed is referred from (Anon n.d.). The fuel consumed and CO2 emitted at study intersections per hour waiting delay is presented in *Table 9* and *10*, respectively.

**Table 9. Fuel consumed by vehicles in idle condition at study intersections**

| Intersection | CNG (kg/h) | Diesel (Lt/h) | Petrol (Lt/h) |
|---|---|---|---|
| SS Circle | 9.42 | 6.56 | 24.63 |
| TH Circle | 5.93 | 8.61 | 11.76 |

**Table 10. CO2 emitted by vehicles in idle conditions at study intersections**

| Intersection | CNG (kg/h) | Diesel (kg/h) | Petrol (kg/h) | Overall (kg/h) |
|---|---|---|---|---|
| SS Circle | 21.21 | 17.32 | 58.91 | 97.45 |
| TH Circle | 13.35 | 22.73 | 28.13 | 64.21 |

The total carbon dioxide emitted is around 160 kg/h during peak time at study intersections. Considering the whole city, the emission is approximately 370 kg/h as there are six signalized intersections. Since the signals are active from 8:00 AM till 9:00 PM, the volume of carbon dioxide released by the automobiles waiting at intersections is ≈4.81 tons/day. There are more than 100 tier-2 cities in India alone; the extent of environmental pollution caused worldwide due to signalized intersections may be anticipated. Primarily the issue can be addressed by optimizing the traffic signal planner to mitigate the waiting delay.

## 8. Conclusion

The primary issues are the urban areas experience several problems concerning road transportation, especially congestion and delay at the signalized intersection. Various probable treatments range from minimum cost measures like signal operation improvement to highest cost measures like junction renovation (FHWA 2004). The choice of treatment depends on the evaluation of the existing operating system. A deep statistical analysis will be conducted to explore the service level of operations at intersections. In this context, a comprehensive study is conducted by selecting Townhall Circle (THC) and Shivakumara Swamiji Circle (SSC) from Tumakuru (tier-2 city), Karnataka state, India. It is observed that the current volume-to-capacity ratios express the stable traffic condition but do not ensure sustainability as the ratios are heading towards congestion. 30% green time wastage is noticed at two intersecting roads of SSC. As per the model, the mean delay is 56 and 36 seconds at SS circle and THC, respectively (Saha et al., 2017). Based on delays, the level of service at SSC and THC is C and B, respectively (Saha et al., 2019).

In comparison, the operation level of both intersections is of class C based on the volume-to-capacity ratio as per HCM (Washington 2005) guidelines. The $CO_2$ emitted during the delay is approximately 370 kg/hr in the city during peak hours. Primary findings of the study are; (a) Significant increase in traffic volume due to intensive use of private vehicles, (b) Probability of congestion in future due to huge traffic volume, and (c) a Large number of emissions are witnessed due to waiting at intersections leading to environmental pollution. The enlisted issues can be addressed through treatments including; (a) Attract commuters towards public transportation, (b) Implementation of vehicle purchase lottery policy as followed in Singapore (Koh 2003) and Beijing (c) Implementation of sustainable and optimized signal operation strategy by applying state-of-art technologies. Finally, the analysis provided in the current article may be used by the city traffic management authorities to identify suitable treatments and implement them accordingly to sustain the stable traffic condition in the city. The same model can be replicated in more than 100 similar tier-2 cities in India. However, it is well known that intersection performance significantly varies on the same corridor. Thus, the proposed study can further be extended with more intersections in the city.

## Acknowledgement

The Smart City Ltd. (SCL) Tumakuru has supported the current work. SCL has shared the traffic data essential for the study. The authors are grateful to the employees of SCL for their cooperation during the study.